# Growth, Properties, and Applications of Pulsed Laser Deposited Nanolaminate Ti$_3$AlC$_2$ Thin Films

Abhijit Biswas[1], Arundhati Sengupta[1], Umashankar Rajput[1], Sachin Kumar Singh[1], Vivek Antad[2], Sk Mujaffar Hossain[1], Swati Parmar[1,3], Dibyata Rout[1], Aparna Deshpande[1], Sunil Nair[1] and Satishchandra Ogale[1,4]*

[1]*Department of Physics and Centre for Energy Science, Indian Institute of Science Education and Research (IISER), Pune, Maharashtra-411008, India*

[2]*Department of Physics, Nowrosjee Wadia College, Pune, Maharashtra 411001, India*

[3]*Department of Technology, Savitribai Phule Pune University, Pune, Maharashtra-411007, India*

[4]*Research Institute for Sustainable Energy (RISE), TCG Centres for Research and Education in Science and Technology (TCG-CREST), Kolkata-700091, India.*

**ABSTARCT**

Recently, nanolaminated ternary carbides have attracted immense interest due to the concomitant presence of both ceramic and metallic properties. Here, we grow nanolaminate Ti3AlC2 thin films by pulsed laser deposition on c-axis-oriented sapphire substrates and, surprisingly, the films are found to be highly oriented along the (103) axis normal to the film plane, rather than the (000l) orientation. Multiple characterization techniques are employed to explore the structural and chemical quality of these films, the electrical and optical properties, and the device functionalities. The 80-nm thick Ti3AlC2 film is highly conducting at room temperature, with a resistivity of about 50 μΩ cm and a very-low-temperature coefficient of resistivity. The ultrathin (2 nm) Ti3AlC2 film has fairly good optical transparency (~70%) and high conductivity (sheet resistance ~735 Ω/sq) at room temperature. Scanning tunneling microscopy reveals the metallic characteristics (finite density of states at the Fermi level) at room temperature. The metal-semiconductor junction of the p-type Ti3AlC2 film and n-Si show the expected rectification (diode) characteristics, in contrast to the ohmic contact behavior in the case of Ti3AlC2/p-Si. A triboelectric-nanogenerator-based touch-sensing device, comprising of the Ti3AlC2 film, shows a very impressive peak-to-peak open-circuit output voltage (~80 V). These observations reveal that pulsed laser deposited Ti3AlC2 thin films have excellent potential for applications in multiple domains, such as bottom electrodes, resistors for high-precision measurements, Schottky diodes, ohmic contacts, fairly transparent ultrathin conductors, and next-generation biomechanical touch sensors for energy harvesting.

## I. INTRODUCTION

Layered nanolaminate ternary carbides and nitrides are a unique class of materials due to the concurrent existence of both ceramic and metallic properties therein; an unusually interesting combination of application-worthy functionalities [1-4]. These ternary materials are popularly known as MAX phases, ($M_{n+1}AX_n$, $n$ = 1, 2 and 3, M is an early transition metal, A is mostly group IIIA and IVA elements, and X is C or N). Until now more than hundred and fifty MAX phase compounds have been discovered, all showing exceptional properties [5].

Among all these ternary carbides and nitrides, $Ti_3AlC_2$ is one of the most studied compounds [6]. Structurally, it has a layered hexagonal symmetry with lattice parameters $a$ = 3.0753 Å and $c$ =18.578 Å, and space group $P6_3/mmc$ [7]. The structure can be envisioned as the alternating two-edge shared octahedral layers of $Ti_6C$ (strong covalent bonding) separated by a two-dimensional (2D) closed packed layer of Al (metallic bonding). It is endowed with excellent thermal and chemical resistance like ceramics, and a good electrical and thermal conductivity. The electrical resistivity of bulk polycrystalline $Ti_3AlC_2$ is ~40 μΩ-cm at room temperature, the temperature coefficient of resistivity (TCR) is small, and as expected, its electronic transport is anisotropic [8-10]. It also shows an unusual combination of properties, e.g. unusual softness (Vickers hardness ~4 GPa), elastic stiffness, chemical robustness, and high damage tolerance at room temperature [1-4]. The discovery of MXene (e.g. $Ti_3C_2$) derived from MAX (e.g. $Ti_3AlC_2$); a 2D analogue of graphene has ushered in a new era of 2D functional carbides and nitrides with enormous functionalities [11-13].

Extensive efforts have been expended by different groups to explore the possibility of obtaining high quality films of different MAX phases (not limited only to the $Ti_3AlC_2$ phase) [3, 14-16]. Pshyk *et al.* reported the growth of $Ti_2AlC$ phase by electron beam physical vapour deposition and studied the nanomechanical and micro-scale tribological properties [17]. Su *et al.* deposited pure-phase polycrystalline $Ti_2AlC$ and $Ti_3AlC$ thin films by reactive radio frequency magnetron sputtering [18]. Rosén *et al.* grew epitaxial $Ti_2AlC$ films by high current pulsed cathodic arc method [19]. Frodelius *et al.* studied the phase stability and initial low temperature oxidation effect in $Ti_2AlC$ thin films grown by magnetic sputtering method [20, 21]. Feng *et al.,* also used magnetron sputtering to grow the $Ti_2AlC$ phase films [22].

Magnuson *et al.* investigated the electronic structure of the dc magnetron sputtered $Ti_3XC_2$ (X = Al, Si and Ge) films by soft x-ray emission spectroscopy [23]. Mauchamp *et al.*, grew epitaxial (000*l*) $Ti_2AlC$ phase thin films by dc magnetron sputtering and showed the anisotropic nature of the electronic transport [24]. Scaabarozi *et al.* and Emmerlich *et al.* showed that magnetron-sputtered thin films of MAX-phase $Ti_3SiC_2$, $Ti_4SiC_3$, $Ti_3GeC_2$, $Ti_2GeC$, and $Ti_2SnC$ are relatively good conductors [25, 26]. There have also been reports on the growth of undoped and doped $Cr_2GeC$ thin films by magnetron sputtering and the anisotropic nature of their electronic transport, mediated through electron-phonon coupling [27, 28]. For more exhaustive literatures, one might look at the in-depth review by Eklund *et al* [3, 14], especially from MAX thin film point of view. Surprisingly, in spite of their unique physical properties, there are only a very few reports on the growth of $Ti_3AlC_2$ thin films of the right (*312*) phase and investigations of their various functional properties [13, 29, 30].

Majority of the thin film growths of $Ti_3AlC_2$ have been performed either by chemical vapor deposition (CVD) or DC magnetron sputtering method [13, 29, 30]. In all those cases, a thin TiC seed layer is used to realize better and more uniform film growth. In terms of thin film growth, the technique of pulsed laser deposition (PLD) which is employed herein, offers some distinct advantages over the CVD or sputtering methods, especially truthful stoichiometry transfer from a single target, and high film density leading to epitaxy (depending on substrate-film lattice parameter compatibility) due to enhance adsorbate surface mobility endowed by high energy radicals in laser generated plasma plume [31]. As stated by Eklund *et al.* all these processes (CVD, sputtering, cathodic arc deposition have their own limitations that sets out the demand of other growth methods for not only $Ti_3AlC_2$ depositions but also various other MAX phase thin films [3, 14]. However, we found only one report about the attempt to grow $Cr_2AlC$ MAX phase thin film by PLD, indicating vast compositional deviation from the $Cr_2AlC$ phase [32]. Therefore, we set out to examine the growth, properties and application-worthiness of thin films of a widely explored MAX phase $Ti_3AlC_2$ using PLD. Given the high standing of the PLD and laser molecular beam epitaxy (MBE) techniques in the field of oxide hetero-epitaxy, oxide electronics and spintronics, we felt that that a success in using the PLD method for growth of high-quality films of MAX phases could translate into new device property regimes integrating oxides with MAX phases.

Amongst the goals of this work, one goal was to go down to the ultrathin limit of thickness and explore the occurrence of any emergent novel effects in such ultrathin films. We indeed find that high quality $Ti_3AlC_2$ thin films (remarkably with high orientation along (103)) can be obtained by PLD and these films are not only highly conducting at room temperature (thus useful as bottom electrodes for numerous layered functional materials), but also have very low temperature coefficient of resistivity (useful for high precision measurements). The ultrathin (2 nm) films show fairly high (70%) optical transparency as well as good electrical conductivity (sheet resistance ~735 Ω/□); making them useful as moderate transparent conductors for some optoelectronic applications. Moreover, these $p$-type $Ti_3AlC_2$ films also render impressive rectifying characteristics with $n$-Si, and ohmic contact with $p$-Si. They also exhibit impressive triboelectric sensing device performance, useful for biomechanical energy harvesting.

## II. EXPERIMETNAL METHODS

The $Ti_3AlC_2$ films were grown by PLD (KrF laser of wavelength 248 nm, pulse width 20 ns). The $Ti_3AlC_2$ target was synthesized by first ball-milling a mixture of Ti, Al and TiC powders (1:1.5:2 molar ratio) at 100 rpm for 24 h in Ar atmosphere and then heating the mixture in a covered alumina crucible in a furnace at 1400 °C for 2 h in Ar atmosphere, as per standard protocols [33]. The polycrystalline single-phase $Ti_3AlC_2$ product (as confirmed by X-ray diffraction (XRD), supporting information **Fig. S1** [34]) as obtained from the furnace was in the form of a hard solid, which was then polished and used as a target. For the depositions, we used $c$-$Al_2O_3$ (0001) sapphire, $n$ or $p$-Si (001) and flexible Kapton substrates, depending on the application/characterization goal. Prior to the depositions, atomically flat surfaces of $c$-$Al_2O_3$ substrates were prepared by annealing at 1000 °C for 1 hour. Films on $c$-$Al_2O_3$ and on Si were grown by using the following growth conditions: growth temperature ~700 °C, laser fluence ~3 J/cm$^2$, target to substrate distance ~40 mm and repetition rate of 10 Hz. All the films were grown in vacuum ambient (~10$^{-6}$ mBar). For the nanogenerator device fabrications, we also used flexible Kapton substrates and grew the films at 400 °C, by keeping all other growth conditions identical.

XRD were performed with a Bruker D2 PHASER X-ray Diffractometer. Atomic force microscopy (AFM) was used for the surface topography analysis and measuring the thickness of the films, using Nanosurf AFM (Switzerland). Raman spectra was recorded with a 2.33 eV (~532 nm) excitation energy laser. X-ray photoelectron spectroscopy (XPS) was done with a K-Alpha X-ray Photoelectron Spectrometer (Thermo-fisher Scientific Instrument, UK). It was carried out in an ultra-high vacuum chamber (2×10$^{-9}$ mBar) by using an Al K$_\alpha$ x-ray source with 6 mA beam current (beam spot size on the sample was ~400 μm). Before recording XPS, *in-situ* ion beam etching (energy 1000 eV) was used for 30s to clean off surface contaminations.

Electrical measurements were performed in standard AC transport four-probe method by using a Quantum Design Physical Property Measurement System (PPMS), within the temperature range of $T$ = 5-300 K. For magnetoresistance measurement, magnetic field was applied along the out-of-plane direction.

Scanning Tunneling Microscopy (STM) was performed using Omicron ultra-high vacuum (UHV) at room temperature, with pressure of $10^{-8}$ mBar. Before imaging, films were in-situ annealed for 60 min in three cycles for 20 min each at 150 °C. The images were processed using image analysis software (SPIP 6.0.9, Image Metrology, Denmark). A tungsten (*W*) tip was used for imaging. The parameters used for the imaging are followings: Bias voltage V = 2 V, set point current I = 100 pA (for 2 nm film); V = 1 V and I = 100 pA (for 80 nm film)

*I-V* characteristics of $Ti_3AlC_2$ films grown on *n* and *p*-type Si (001) substrates were measured using Keithley 4200 Semiconductor Characterization System (SCS) connected to a standard probe station (Semiprobe, USA). Depending on the position of the tungsten (W) probe on the devices, two modes were defined (forward bias and reverse bias). All the *I-V* measurements were carried out in current perpendicular to plane (CPP) mode with voltage sweep of ±3V having a voltage step of 0.05V at $T$ = 300K.

Triboelectric nanogenerator (TENG) devices were fabricated using $Ti_3AlC_2$ films. A contact mode TENG architecture was adopted for the design of the touch sensor. To maintain the gap between the top and bottom layer, four 3M double side tapes (1 mm thick) were also used, and were attached at the edges. Total active device area was 1.5×1.5 cm$^2$. The electrical outputs were measured using a Keithley DMM7510 multimeter by soldering copper wires on the copper tapes on both electrodes (Al and $Ti_3AlC_2$), with short-circuit current of 3 μA. We used the PASCO Force sensor to measure the applied force from human-arms, is of ~15 N.

**III. RESULTS AND DISCUSSION**

**A. Structural characterizations**

The X-ray diffraction (XRD) patterns shown in **Fig. 1(a)** indicate the presence of only two film peaks in addition to the (000*l*) substrate peaks corresponding to the *c*-axis sapphire. We had expected that the (000*l*) orientation of the substrate would translate into the film and as such, we looked for the same in the XRD. However, in this case one must see multiple peaks corresponding to (000*l*, *l* = 2, 4, 6 etc.), starting from (0002) 2θ≈9.44°. However, surprisingly, we only found only two peaks near (0008) and (00016) and no other peaks. Thus, it became clear that the film does not grow along the nominally expected (000$l$) orientation. Interestingly, the XRD peaks for the (103) and (206) planes of $Ti_3AlC_2$ are also quite close to the same two 2θ value where the clear but somewhat broader peaks are seen **Fig. 1(b)**. The led us to conclude that the film grows in the form of high degree of orientation along the (103) orientation normal to the film plane.

The interesting choice of (103) orientation by the growing film in the PLD process can be partly elucidated by the comparison of the possible surface terminations of the $Ti_3AlC_2$ lattice surface along the (000$l$) and (103) orientations and related possible interface chemistry scenario. Towards this end, we present in **Fig. 1(c)** the surface atomic arrangements in the cases of the two orientations with top and side views. It is clear that in the case of the (000$l$) orientation, the surface terminations are either in the form of the Ti or Al or C planes, while the terminations for the (103) orientation, are mixed atom terminations. In the single $Ti_3AlC_2$ target based PLD process, all the atoms from the target (Ti, Al, C) and corresponding radicals impinging on the hot (700 °C) substrate at the same time. The atoms (with energy in the range of 0.1 eV to several eV for the pulsed laser generated plasma) undergo physisorption and surface diffusion followed by chemisorption or incorporation into the growth front. Initially clusters get formed and they grow laterally and vertically as driven by the in-plane and out-of-plane step diffusion rates that are co-guided by the growth temperature and the chemistry of bonding on the surface and at the step edges. Since the top of the $c$-axis oriented sapphire substrate would basically have fully oxygenated surface, the surface oxygens would strive to bond with the impinging Ti, Al, C atoms and the corresponding single atom or multi-atom compound radicals. Since the Ti, Al and C atoms arrive in the nominal proportion of about 3:1:2, the chances of forming initial Ti-C related local clusters with scattered presence of low density Al atoms are higher than the formation of facile stoichiometric Ti-Al-C clusters. Although the detailed analysis will have to follow complex Monte Carlo growth simulation, in the absence of it one could argue that the above scenario does not quite favor the formation of successive ordered layers of Ti, Al and C that will have to form for the film to acquire (000$l$) orientation. It would need selective upward (and downward) step diffusion of one specie over the other, which could be kinetically more complicated rather than a scenario wherein all three atom types are available in the same plane and rearranging themselves laterally more than vertically. Given the fact that the lattice symmetry and parameters for the two plane orientations (as shown in **Fig. 1(c)**) are quite similar vis a vis the sapphire top surface for $c$-axis orientation, it appears that the film chooses (103) orientation at least under the growth conditions availed in this work. Importantly, as shown later, the film stoichiometry does truthfully follow the target stoichiometry as is known for the PLD process, namely Ti:Al:C as 3:2:1. Moreover no other secondary XRD peaks are seen to suggest multiple phases. Thus, (103) oriented growth of $Ti_3AlC_2$ is strongly indicated.

The forgoing arguments do not mean that there could not be another PLD parameter space which could lead to (000*l*) film orientation. One clue to the non-triviality in this respect appears from the broad nature of the (103) and (206) peaks, which suggest presence of some interface diffusion and reaction leading to tilt angle distribution of grain orientation. Most appropriate strategy under this condition would possibly be the introduction of a diffusion barrier layer such as the TiC (111) layer insightfully employed in some works on (000*l*) oriented film growth [13, 29, 30]. Indeed, it has been suggested that such a layer minimizes the high in-plane strain and also improves the wetting [13, 29, 30]. We will continue to pursue the work on different barrier layers in further studies, because the corresponding parameters for all-PLD process will have to be differently optimized as compared to the other techniques such as sputtering.

The surface topography revealed by atomic force microscopy (AFM) mimics a clear step-terrace pattern of the substrate for ultrathin films. Indeed, all the films have atomically smooth surfaces with surface roughness of less than one nm (inset of **Fig. 1(a)** and **Fig. S2**).

We also examined the elemental ratio of $Ti_3AlC_2$ film (deliberately grown on Si to avoid non-separable contribution of Al from $Al_2O_3$ substrate) by Field Emission Scanning Electron Microscopy (FESEM) - energy-dispersive X-ray spectroscopy (EDAX) analysis, and it showed the Ti:Al ratio close to ~3:1 (35.7:11.3) as expected (**Fig. S3a**). This confirms the truthful stoichiometry transfer by the PLD process, which is its hallmark. Furthermore, in order to verify the formation of hexagonal $Ti_3AlC_2$ phase, we made an interesting study by depositing $Ti_3AlC_2$ directly on carbon coated Cu grid by PLD (at room temperature) and recorded the top view high resolution transmission electron microscopy (HRTEM) image. This can gives us some indication about the initial stage of thin film growth. From the image, we can clearly see the local formation of hexagonal lattice structure (**Fig. S3b, Fig 1c**). The EDAX once again confirmed the elemental ratio of Ti and Al is ~3:1 (76.1:23.9); further confirming that the film grown by PLD is hexagonal $Ti_3AlC_2$.

For gaining further insights into the microstructural and bonding aspects, we recorded room temperature Raman spectra for the case of bulk 80 nm film. For the $Ti_3AlC_2$ phase, based on the symmetry considerations, ideally we should see seven Raman active modes ($2A_{1g}$ +

$2E_{1g} + 3E_{2g}$) [35], where $A_{1g}$ and $E_{1g}$ correspond to the out-of-plane vibrations, and $E_{2g}$ corresponds to the in-plane vibration, respectively. However, literature reports have shown that only six modes are visible in the spectra [36], which we clearly see for our film as well (**Fig. 2(a)**). It is to be noted however that the main peak in the range of 600-700 cm$^{-1}$ is considerably broad [36], possibly suggesting multiple contributions (such as intrinsic inhomogeneity related with the chemical compositions, grain size and morphology) as well as a degree of expected strain gradient (distribution) in the film. We show the fitting of the broad signature with multiple contributions (inset of **Fig. 2(a)**). Two of these ($A_{1g}$ and $E_{2g}$) correspond to the pure $Ti_3AlC_2$ phase [36], albeit with a small strain induced shift. The other relatively smaller contributions emanate from the surface oxide known to be in the form of $Ti_2CO_2$ [11] and possibly small contributions from the sapphire substrate. Additionally, we also observed a Raman signature around 400 cm$^{-1}$ (peak indicated by an asterisk), as reported also by Presser *et al,* associated possibly with a degree of disorder [36]. These data reflect that $Ti_3AlC_2$ film has a thin layer of surface oxide because of the known high reactivity of the surface.

We also performed X-ray photoelectron spectroscopy (XPS) on the in-situ mildly sputter-cleaned 80 nm $Ti_3AlC_2$ film. The XPS data for the Ti 2p, Al 2p and C 1s are shown **Fig. 2(b)-(d)**. Peak fittings were performed using CasaXPS software. As noted by other groups, the Ti-2p XPS data for the as-grown film (**Fig. S4**) shows multiple peaks due to the surface oxidation [37]. The C-1s XPS for as-received films also shows the presence of carboxylate (-COO) surface groups [13]. Notably, after in-situ cleaning, the spectrum changes drastically, especially for Ti-2p and C-1s. The Ti-2p spectrum can be best fitted with three pairs (for $2p_{1/2}$ and $2p_{3/2}$) of curves with the universal spin-orbit peak splitting of ~6 eV (**Fig. 2(b)**). These three pairs can be assigned to Ti-Al-C, Ti (IV) oxide and Ti (III) oxide, respectively. The spectrum matches quite well with the reported data of CVD grown $Ti_3AlC_2$ thin film [13].

Interestingly, in case of Al-2p we could only see the higher binding energy peak of Al oxide (**Fig. 2(c)**) and did not observe the expected metallic Ti-Al peak at ~72 eV. We believe that the high reactivity of $Ti_3AlC_2$ causes oxidation of the top few layers where the major XPS signal emanates from, and therefore the feature is either small or negligible. The absence of Ti-Al peak was also reported for ceramic $Mo_2TiAlC_2$ sample [38]. In our case, wherein the deposition is performed directly on sapphire without a nucleating (or barrier) layer of TiC as done in several other papers [13, 29], the high substrate induced strain leads to the film growth by 3D island or grain growth mode (as also confirmed by AFM) rather than a layer-by-layer mode. Thus, in addition to the bulk diffusion the columnar morphology will cause grain boundary oxygen diffusion leading to higher degree of oxidation of the top few layers. Eklund *et al*., have clearly stated that when growing a film of $Al_2O_3$ at high temperature, Al and O migrates from the substrate, and at the film–substrate interface substrate species also form along with the deposited species [14]. Interestingly we found out that the Al-2p spectrum does show the expected ~72 eV contribution from Ti-Al when the film was grown at much lower growth temperature of 400 °C (**Fig. S5**). This implies that the oxidation most possibly occurs during growth at high temperature due to residual oxygen present in the system.

For the C-1s core, the peak at the low binding energy corresponds to Ti-C (**Fig. 2(d)**) [13, 37]. In addition, importantly, the contribution from Ti-C bond enhances significantly after in-situ cleaning of the surface (**Fig. 2(d)** and **Fig. S4**), reflected by the absence of carboxylate (-COO) surface groups.

**B. Carrier transport**

The electrical measurement on the 80 nm thick $Ti_3AlC_2$ film shows resistivity ($\rho$) ~50 µΩ-cm at $T = 300$ K (**Fig. 3(a)**), comparable to the bulk polycrystalline $Ti_3AlC_2$ phase (~40 µΩ-cm at $T = 300$ K) [8-10, 39]. This low $\rho$ at $T = 300$ K is quite remarkable, a consequence of the presence of substantial density of states (DOS) at the Fermi level dominated by the *d-d* orbitals of the Ti [7]. Interestingly, in our $Ti_3AlC_2$ films the derivative of resistivity ($d\rho/dT$) is very weakly *T*-dependent (inset of **Fig. 3(a)**) with resistivity value of $\rho$ ~62 µΩ-cm, even at $T = 5$ K. The temperature coefficient resistivity (TCR) (defined as $\alpha = \frac{1}{\rho_T}\left(\frac{d\rho}{dT}\right)$, $\rho_T$ at a given temperature) is found to be ~4.5 ×$10^{-4}$ $K^{-1}$ (~360 ppm/°C) within a temperature range $150 \leq T \leq 200$ K; characteristics of a bad metal where quasiparticle states are collapsed due to the strong inelastic scattering [40].

Hall Effect measurement at $T = 300$ K shows (data not included here) that charge carriers are of *p*-type (holes are the majority carriers) with typical metal-like carrier concentration ($n_p$) of ~$10^{28}/m^3$ and mobility ($\mu$) ~$7.5 \times 10^{-4}$ $m^2V^{-1}s^{-1}$, comparable to the observed values for the bulk polycrystalline $Ti_3AlC_2$ [8]. Magnetoresistance (MR), defined as $\frac{\Delta\rho}{\rho_0} = \frac{\rho(B)-\rho(0)}{\rho(0)}$), is noted to be positive with quadratic magnetic field dependence (MR $\propto B^2$) (**Fig. 3(b)**), as is usually observed in a typical metal, as a consequence of the Lorentz contribution. All these analyses indicate that at room temperature, our PLD grown layered $Ti_3AlC_2$ thin film is highly conducting with holes (*p*-type) as majority charge carriers, and electronically it lies on the boundary between a metal and semiconductor.

We further examined the thickness dependent sheet resistance ($R_S$) and optical transmission by examining films from 80 nm thick down to ultrathin 2 nm, since ultrathin films can potentially generate interesting emergent phenomena as a consequence of quantum confinement, high surface to volume ratio, surface states/charge, and interfacial strain. Upon reducing the film thickness, the $R_S$ was noted to increase (**Fig. 4 and S6**) due to the reduction of band filling as well as the increase of effective disorder (defined as disorder/band width), as the disorder becomes more effective in scattering of charge carriers in thinner films. Surface and interface scattering contributions as well as grain boundary scattering contributions are known to be accentuated with reduction in film thickness below a certain value commensurate with the mean free path [41]. Interestingly, even the ultrathin 2 nm film shows an impressive and application-worthy conductivity ($R_s$ ~735 Ω/□) at $T$ = 300 K. Equally important, the TCR is found to be very low (as the $dR_S/dT$ is very low in **Fig. S7**), confirming that the basic material character of being on the metal-semiconductor boundary and its properties are retained even in few unit-cell ultrathin films.

Understanding the nature of electronic transport in $Ti_3AlC_2$ phases is quite tricky as revealed by several interesting past studies. Barsoum *et al.* have reported on the temperature dependence of resistivity of the polycrystalline samples of two 312 ($Ti_3AlC_2$, $Ti_3SiC_2$) and one 413 ($Ti_4AlN_3$) MAX phases over the wide temperature regime and interpreted (compared) the data in terms of the charge carrier concentrations and defect scattering contributions [8]. They argued that the conductivity is lower in $Ti_3AlC_2$ due to dearth of carriers. A detailed analysis by Finkel *et al.* [9] again on polycrystalline $Ti_3AlC_2$ emphasized the need of invoking two-band framework to explain the transport for the case of $Ti_3AlC_2$ and suggested that carrier concentrations remain almost constant over broad temperature range while mobility changes due to phonon scattering and neutral impurities. Reports on $Ti_2AlC$, $Cr_2GeC$ thin films suggest that electron-phonon interaction also plays an important role for the electronic transport [24, 27]. It is to be noted that the room-temperature resistivity of $Ti_3AlC_2$ film (grown by magnetron sputtering) found in the literature is ~40-50 μΩ-cm [8, 9, 29], similar as our PLD grown films.

In our case, for the 80 nm film, except for a small drop in resistivity with decreasing temperature over the 270 ≤$T$≤ 300 K range, the resistivity is seen to increase very mildly with decreasing temperature. Although the room temperature resistivity of our film is very close to the reported one, it shows a mild increase in resistivity while cooling the temperature, an opposite trend with respect to the normal metal-like characteristics. Notably the rise in resistivity with lowing temperature is not very rapid, as expected for a semiconductor either.

This can happen if the film has a lateral grain-like structure (even though it is oriented) and the grain boundary transport competes with intra-grain metallic transport. Therefore, one can argue that the observed behavior is a consequence of the competition between the inter-grain transport (at the grain boundaries) and de-localized intra-grain metal-like transport.

The grain boundary aspects in polycrystalline samples and their possible contributions to transport have also not yet been fully addressed. Therefore, one can argue that possibly moderate strain gradient disorder or density of point defects due to possible slight off-stoichiometry could cause localization effects rendering a mild upturn of the resistivity. Since the suggested two-band model envisages the active role of both carrier types, preferential localization of one over the other can affect the overall majority/minority carrier concentration as well as their respective mobilities. Interestingly, the MR observed in our case at $T=$ 5K is not only positive but has the value of the quadratic coefficient $\alpha = 1.03 \times 10^{-2}$ m$^4$/V$^2$s$^2$ (MR = $\alpha \times B^2$), which is much higher than that reported value of $\sim 10^{-4}$ m$^4$/V$^2$s$^2$ at $T=$ 5K [9]. A small concentration of defects or disorder can render hopping behavior and this can be influenced by the magnetic field leading to the noted positive MR.

## C. Optical transmission

Interestingly, while the bulk 80 nm film is expectedly black due to high conductivity, the ultrathin (2 nm) film is fairly transparent (~70%) in the visible range in spite of its high conductivity. As expected, the transparency decreases systematically with the increase of the film thickness (**Fig. 5a and S8**). In our case, we have noted that 4 nm Ti$_3$AlC$_2$ film is ~55% transparent, whereas 8 nm film is ~30 % transparent and the 80 nm one is black (**Fig. 5b and Table 1**). Considering the highly conducting nature of Ti$_3$AlC$_2$ films with their ceramic-type robustness (high scratch resistance) they could possibly be used in some applications in the place of wide-band gap semiconductor films (e.g. tin-doped indium oxide (ITO)) [42].

## D. Scanning tunnelling microscopy (STM)

We also performed in-situ scanning tunneling microscopy (STM) for both the ultrathin (~2 nm) as well as thicker Ti$_3$AlC$_2$ (~80 nm) films at $T$ = 300 K to elucidate the density of electronic states. To the best of our knowledge, this is the first STM evaluation on Ti$_3$AlC$_2$ phase thin films. STM topographic images of the in-situ annealed 2 nm and 80 nm Ti$_3$AlC$_2$ films are shown in **Fig. 6(a) and (b)**. To gain insight into the nature of the density of electronic states, we also measured the *I-V* characteristics. Numerical differentiation of *I-V* curves yielded dI/dV spectra, which are a measure of the local density of states (LDOS).

The STM dI/dV data for the as-grown sample exposed to air and measured in ambient conditions shows specific features typical of localized states, which can be attributed to the surface groups (**Fig. S9**). However, for the in-situ annealed films (and the subsequent STM recorded inside the UHV chamber), we note a parabolic-type (not-strictly parabolic) nature of differential conductance with a finite DOS at Fermi level ($E_F$), signifying metallicity for the in-situ annealed films (**Fig. 6c and d**). Interestingly, around the $E_F$, the nature of this dependence shows resemblance with the calculated DOS by Zhou *et al.* which mainly originates from the nearly free-electron state of Ti-$3d$ [7]. It may be noted that although the electronic structure calculations about this material are interesting, the experimental work is lacking on same, with subsequent fabrication and evaluation of device functionalities. Our work is a step in this direction, since PLD process gives immense control on the surface quality and subsequent properties of the films and their in-situ hetero-structures (known well in the field of oxides).

**E. Schottky diodes**

Considering the highly conducting yet robust ceramic nature of $Ti_3AlC_2$, (unlike the noble metals such as Au and Ag which are mechanically soft with issues of thermal stability and high current induced materials transport in the form of electro-migration), we examined the electrical (*I-V*) characteristics of the heterojunction between *p*-$Ti_3AlC_2$ film and *n*-Si (semiconductor) substrate, a Schottky diode configuration. An ultrathin layer of surface oxide, revealed by XPS, in principle may qualify it more as a *p-i-n* tunnel junction rather than a pure *p-n* junction. The *p*-$Ti_3AlC_2$/*n*-Si junction exhibited a clear rectifying character (**Fig. 7a**), consistent with the *p*-type charge carriers of the $Ti_3AlC_2$ as discussed earlier. Significantly low forward junction potential reveals the typical nature expected for a Schottky junction. The forward to reverse current ratio at 0.5 V (-0.5 V) is 21.3, while at 1 V (-1V) it is 36.6. In contrast, for the *p*-$Ti_3AlC_2$/*p*-Si junction, we observed a non-Schottky behavior in the *I-V* characteristic (**Fig. 7b**), as both layers are of *p*-type, but with a degree of non-linearity. Recently, Fashandi *et al.*, have shown the formation of Ohmic contact for various 312 MAX phases grown on SiC with remarkable stability even after 1000 hrs of ageing at 600 °C in air [43]. Moreover, rapid thermal annealing (900 °C for 4 min) of $Ti_3SiC_2$ also results in formation of Ohmic contact [44]. We also annealed the *p*-$Ti_3AlC_2$/*p*-Si film (in-situ annealing inside the PLD chamber at 700 °C for 3 hrs following deposition) to check whether the *I-V* becomes fully Ohmic or not. Interestingly, the *I-V* curve show enhanced linearity (**Fig. S10**) with the formation of almost Ohmic contact (within the depicted small range of applied voltage) for Ti$_3$AlC$_2$/*p*-Si (inset of **Fig. 7b**). The transport across the Schottky contact (*p-n* junction) can be viewed within the framework of the thermionic emission theory, diffusion theory, or the effect of both depending upon the specific theoretical models [45]. Since these are related to the mobility, specifically high mobility (in case of thermionic emission) and low mobility (in case of diffusion), we conjecture that diffusion across the interface is mostly applicable in our case as Ti$_3$AlC$_2$ shows rather low mobility at *T* = 300 K ($\mu$ ~10$^{-4}$ m$^2$V$^{-1}$s$^{-1}$). This metal-semiconductor junction feature can be attributed with the Fermi level ($E_F$) pinning by the electronic mid-gap states or the interface defect states, as nanoscale inhomogeneity is ubiquitous at the interface of thin film heterostructures [45].

**F. Triboelectric Nano-generator (TENG)**

Finally, having confirmed the highly metallic nature with *p*-type charge carrier of Ti$_3$AlC$_2$ films, we proceeded to explore the efficacy of these films as an electron accepting electrode in a the triboelectric nanogenerator (TENG) application. In general, TENG devices are in high demand for sensing, tribotronics or mechanical energy harvesting applications [46, 47]. Indeed, energy harvesting from human movements for portable and wearable electronic devices has received considerable attention these days. However, the materials currently in use are limited to a few polymeric systems (including some wherein active materials are embedded in polymers thereby forming composites) wherein issues of efficient charge transport and transfer, as well as bond-cleaving, wear and physical transfer of material are matters of continuing concern [48]. Recently, several novel designs of TENGs including use of 2D materials and biomaterials have been researched with very impressive outcomes [49]. Since MXene has laterally slidable layers with low out-of-plane conductivity [12], we conjectured that even thin films of the robust Ti$_3$AlC$_2$ phase with excellent mechanical stability and high conductivity could render interesting TENG effects. Thus, we fabricated and tested TENG devices by using Ti$_3$AlC$_2$ films grown on flexible Kapton substrate, which enables homologous interface contact between the actuator and the device. This also suits the TENG requirement to adapt to flexible/wearable devices.

The schematic of a typical TENG device is shown (**Fig. 8a**) where *p*-type conducting Ti$_3$AlC$_2$ was used as a part of the touch sensor device. As stated above, the Ti$_3$AlC$_2$ does not only provides +ve charges (p-type), but also acts as a highly conducting bottom electrode for the contact. In contrast, glass microfiber filter paper was used to supply –ve charge carriers. On the top of it, aluminium was used as a top electrode. Functionally, when both the layers are almost in contact with each other (by pressing), both *p* and *n*-type carriers feel dipolar forces and exchange of carriers takes place at the interface; generating triboelectric surface charges at the interface. When the pressure is released, the charges (both +ve and –ve) move away from each other. However, to compensate the interfacial triboelectric effect, they build up opposite charges on the bottom surface (of $Ti_3AlC_2$) and top surface (of aluminium), generating a potential difference. Our TENG device showed impressive peak-to-peak open circuit output voltage ($V_{oc}$) ~80 V (**Fig. 8b**) which could be attributed to efficient contact electrification that can be enhanced by the interface texture of the $Ti_3AlC_2$ grown on Kapton.

## IV. CONCLUSIONS

In summary, layered nanolaminate $Ti_3AlC_2$ thin films are grown by the pulsed laser deposition (PLD) technique (more known in the field of metals oxides, oxide electronics/spintronics). The films show a truthful transfer of the stoichiometry from the target to the film, a hallmark of the single target PLD technique. Remarkably, the film are noted to grow with (103) orientation normal to the film plane. These films are shown to exhibit various interesting application-worthy physical properties, especially high conductivity with low TCR even down to very low temperature. They are fairly transparent in ultrathin form yet retaining very high conductivity at room temperature. Furthermore, for the first time scanning tunneling microscopy (STM) was employed to characterize the $Ti_3AlC_2$ films, shows metallic-characteristics at room temperature. The PLD grown $Ti_3AlC_2$ also shows rectification (diode) characteristics (*p-n* junction) with *n*-Si and Ohmic-like contact (*p-p* junction) with *p*-Si. The films also show a remarkable TENG-based biomechanical touch-sensing capability. The ability to grow high quality MAX phase thin films by PLD can open up interesting avenues for their integration with functional metal oxide to generate novel properties [50].

**ACKNOWLEDGMENTS**

S. B. Ogale would like to acknowledge funding support by the Department of Atomic Energy, via the Rama Ramanna Fellowship and the funding support of DST Nanomission Thematic unit (SR/NM/TP-13/2016). A. Biswas would like to thank SERB-N-PDF fellowship (PDF/2017/000313). A. Sengupta would like to thank SERB-N-PDF fellowship (PDF/2017/001957). V. Antad is thankful to DST-SERB for the project funded under SERB-TARE-2018 scheme (TAR/2018/000807). We would like to thank Dr. Surjeet Singh, Varun Natu, Rajesh Mandal, Soumendu Roy, Anupam Singh, Rushikesh Magdum for the help in various measurements and analysis.

# REFERENCES

[1] M. Radovic and M. W. Barsoum, **MAX phases: Bridging the gap between metals and ceramics**, Am. Ceram. Soc. Bull. **92**, 20 (2013).

[2] M. W. Barsoum, **The $M_{N+1}AX_N$ Phases: A New Class of Solids**, Prog. Solid St. Chem. **28**, 201 (2000).

[3] P. Eklund, M. Beckers, U. Jansson, H. Högberg and L. Hultman, **The $M_{n+1}AX_n$ phases: Materials science and thin-film processing,** Thin Solid Films, **518**, 1851 (2010).

[4] M. W. Barsoum and M. Radovic, **Elastic and Mechanical Properties of the MAX Phases**, Ann. Rev. Mater. Res. **41**, 195 (2011).

[5] M. Sokol, V. Natu, S. Kota and M. W. Barsoum, **On the Chemical Diversity of the MAX Phases**, Trends in Chemistry, **1**, 210 (2019).

[6] N. V. Tzenov and M. W. Barsoum, **Synthesis and Characterization of $Ti_3AlC_2$**, J. Am. Ceram. Soc. **83**, 825 (2000).

[7] Y. C. Zhou, X. H. Wang, Z. M. Sun and S. Q. Chen, **Electronic and structural properties of the layered ternary carbide $Ti_3AlC_2$**, J. Mater. Chem. **11**, 2335 (2001).

[8] M. W. Barsoum, H. –I. Yoo, I. K. Polushina, V. Yu. Rud, Yu. V. Rud and T. El-Raghy, **Electrical conductivity, thermopower, and Hall effect of $Ti_3AlC_2$, $Ti_4AlN_3$ and $Ti_3SiC_2$**, Phys. Rev. B **62**, 10194 (2000).

[9] P. Finkel, M. W. Barsoum, J. D. Hettinger, S. E. Lofland and H. I. Yoo, **Low-temperature transport properties of nanolaminate $Ti_3AlC_2$ and $Ti_4AlN_3$**, Phys. Rev. B **67**, 235108 (2003).

[10] G. Ya Khadzhai, R. V. Vovk, T. A. Prichna, E. S. Geovorkyan, M. V. Kislitsa and A. L. Solovjov, **Electrical and thermal conductivity of the $Ti_3AlC_2$ MAX phase at low temperatures**, Low Temp. Phys. **44**, 451 (2018).

[11] M. Naguib, M. Kurtoglu, V. Presser, J. Lu, J. Niu, M Heon, L. Hultman, Y. Gogotsi and M. W. Barsoum, **Two-Dimensional Nanocrystals Produced by Exfoliation of $Ti_3AlC_2$**, Adv. Mater. **23**, 4248 (2011).

[12] M. Naguib, V. N. Mochalin, M. W. Barsoum and Y. Gogotsi, **25th Anniversary Article: Mxenes: A New Family of Two-Diemnsional Materials**, Avd. Mater. **26**, 992 (2014).

[13] J. Halim, M. R. Lukatskaya, K. M. Cook, J. Lu, C. R. Smith, L. Näslund, S. J. May, L. Hultman, Y. Gogotsi, P. Eklund and M. W. Barsoum, **Transparent Conductive Two-**
16

# TABLE

**TABLE 1**. Transparency and room temperature sheet resistance ($R_S$) of $Ti_3AlC_2$ films grown on *c*-$Al_2O_3$ (0001) substrates.

| Film | Thickness (nm) | Transparency (%) | Sheet Resistance ($R_S$ ~Ω/□) |
|---|---|---|---|
| $Ti_3AlC_2$ | 2 | ~70 | ~735 |
| | 4 | ~55 | ~580 |
| | 8 | ~30 | ~360 |
| | 80 | ~8 | ~20 |

**FIGURE CAPTIONS**

**FIG. 1. (Color online)** (a) X-ray diffraction pattern of $Ti_3AlC_2$ thin films of various thicknesses grown on $c$-$Al_2O_3$ (0001) substrates (broad humps at lower angle is due to the sample holder, as shown) Inset shows topographic image by atomic force microscopy of a $Ti_3AlC_2$ film (2 nm) showing atomically smooth surface with clear step-terrace structure and roughness of 0.377 nm. (b) In XRD, separate contributions emanating from substrate (S) and film (F) is clearly visible. (c) Schematic of the $Al_2O_3$ (000$l$) and $Ti_3AlC_2$ along (000$l$) and (103) planes with their top-view (left column) and side-view (right column).

**FIG. 2. (Color online)** (a) Raman spectrum and (b)-(d) X-ray photoelectron spectra of in-situ surface cleaned 80 nm $Ti_3AlC_2$ film. Asterisk in Raman spectrum corresponds to the possible disorder as well as substrate related peak.

**FIG. 3. (Color online)** Temperature dependent carrier transport of 80 nm $Ti_3AlC_2$ film grown on $c$-$Al_2O_3$ (0001) substrate. (a) Resistivity shows highly conducting nature at room temperature with very small variation in resistivity over the whole temperature range (inset). (b) Positive quadratic field dependent magnetoresistance, showing normal metal-like behavior.

**FIG. 4. (Color online)** Thickness dependent sheet resistance ($R_S$) of $Ti_3AlC_2$ thin films grown on $c$-$Al_2O_3$ (0001) substrates.

**FIG. 5. (Color online)** (a) Transparency of both-side polished bare $c$-$Al_2O_3$ (0001) substrate (left) and deposited 2 nm $Ti_3AlC_2$ ultra-thin film (right). (b) Thickness dependent optical transmission in the visible wavelength range shows that ultrathin 2 nm film transmits ~70% of the light, having sheet resistance of ~735 $\Omega/\square$ at room temperature.

**FIG. 6. (Color online)** (a)-(b) In-situ Scanning tunnelling microscopy (STM) topographic images, (c) current vs. voltage and (d) differential conductance (dI/dV) spectra of 2 nm and 80 nm $Ti_3AlC_2$ films taken at room temperature within the applied bias voltage range of ±1V.

**FIG. 7. (Color online)** (a) Room temperature *I-V* characteristics of 80 nm $Ti_3AlC_2$ film grown on *n*-Si (001) substrate showing rectification behavior. (b) In contrast, almost Ohmic-like contact was observed for film grown on *p*-Si (001). After in-situ annealing at 700 °C for 3 hrs, in small bias voltage range, fully Ohmic contact was observed in case of *p*-$Ti_3AlC_2$/*p*-Si, in small bias range (inset).

**FIG. 8. (Color online)** (a) Schematic of a triboelectric nanogenerator (TENG) device containing $Ti_3AlC_2$ film grown on flexible Kapton substrate. (b) Open circuit voltage ($V_{oc}$) of the TENG based touch sensor device.

**(FIGURE-1)**

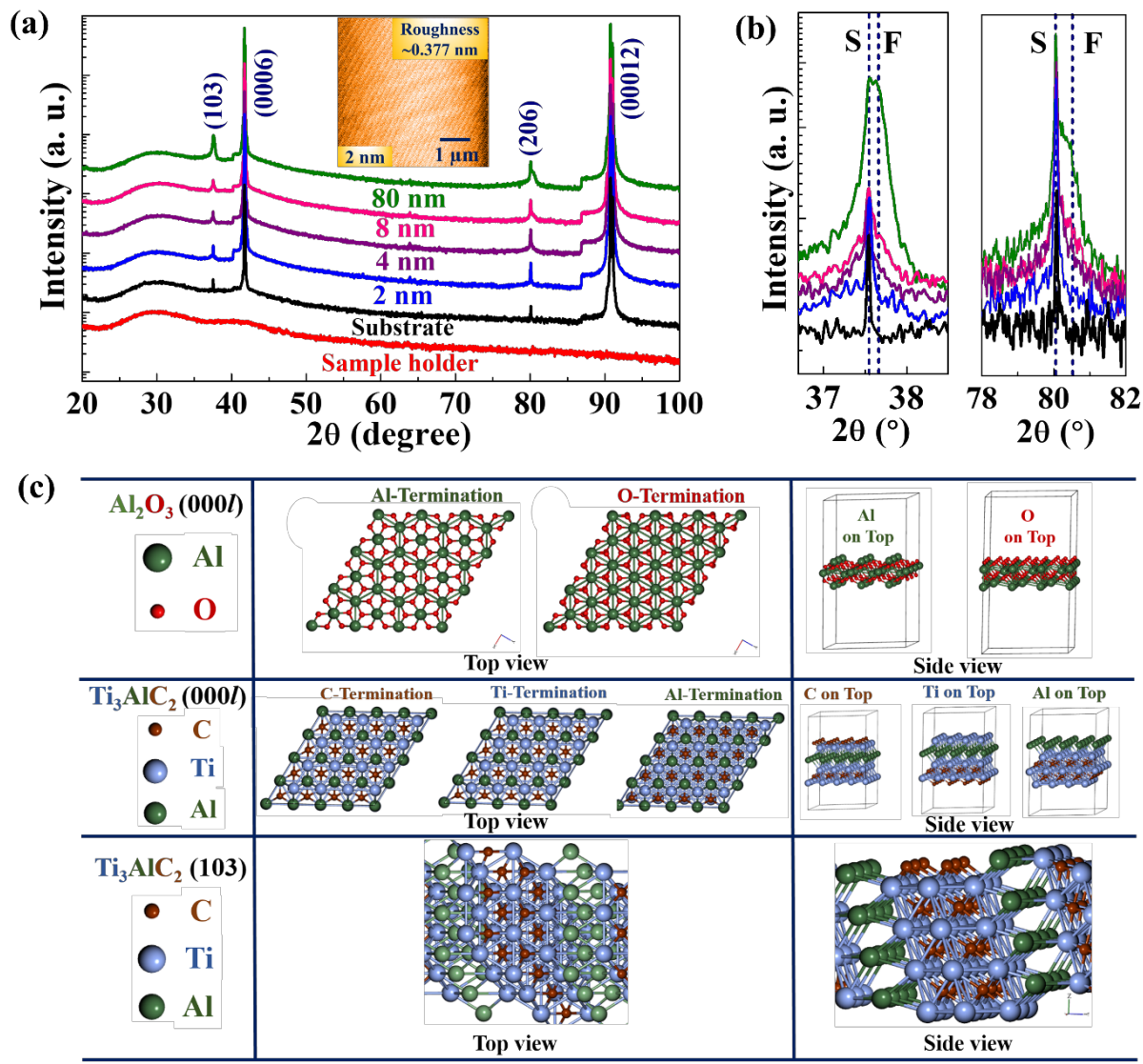

**(FIGURE-2)**

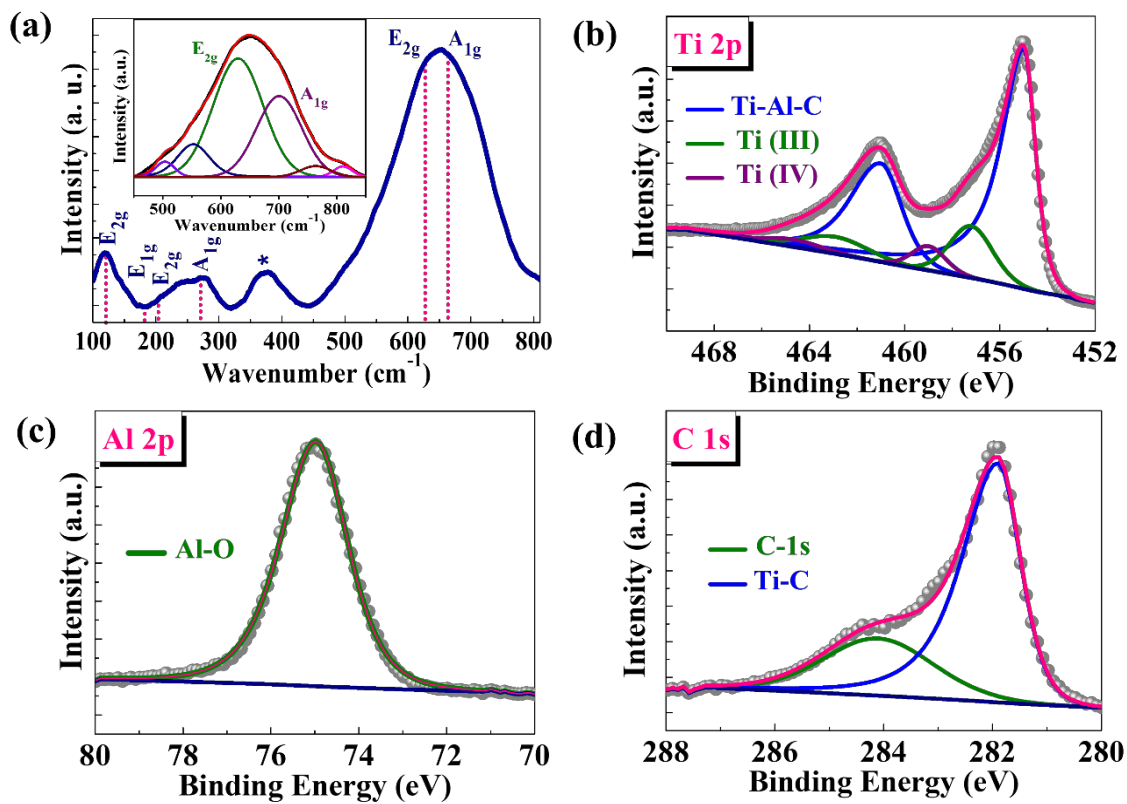

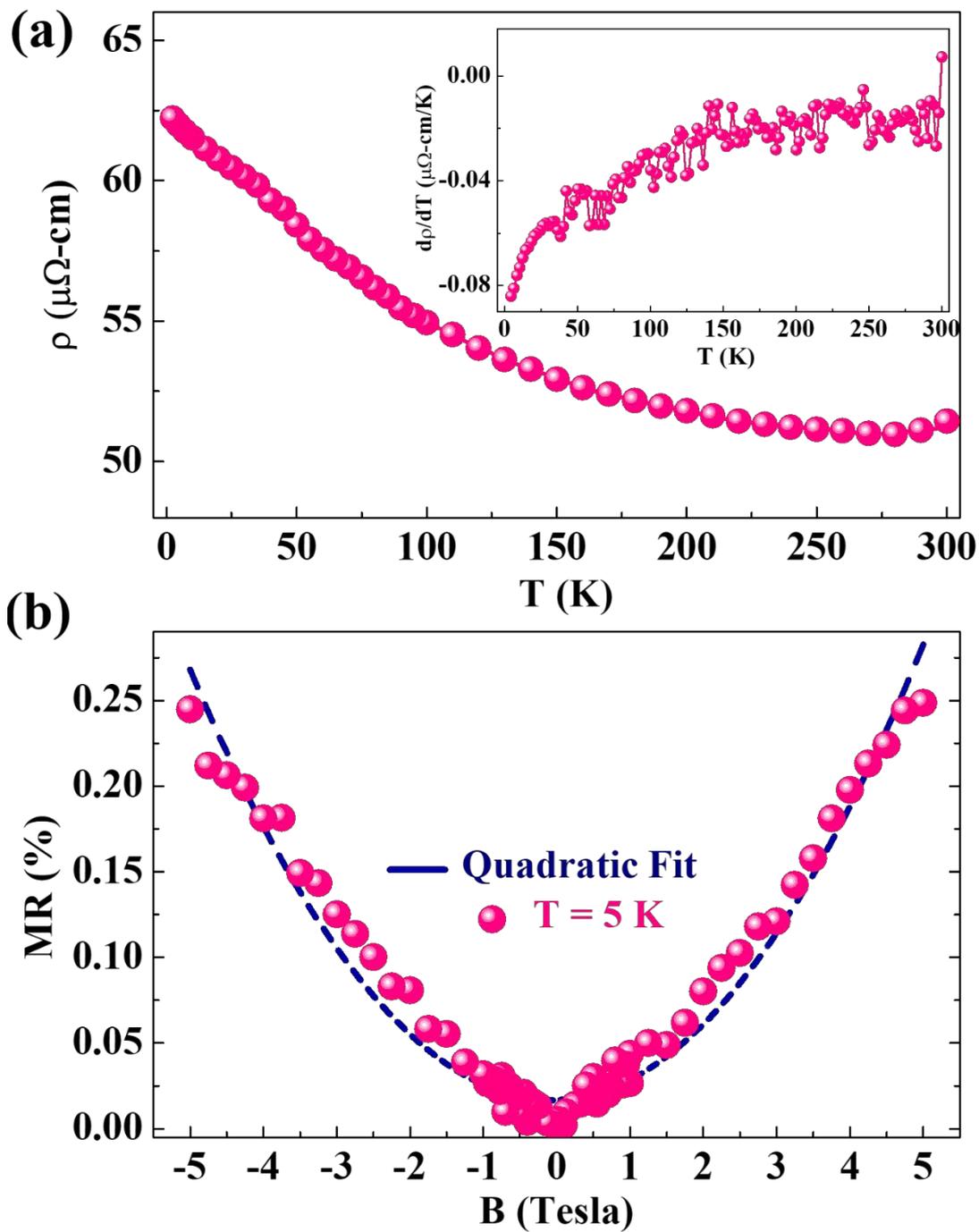

**(FIGURE-4)**

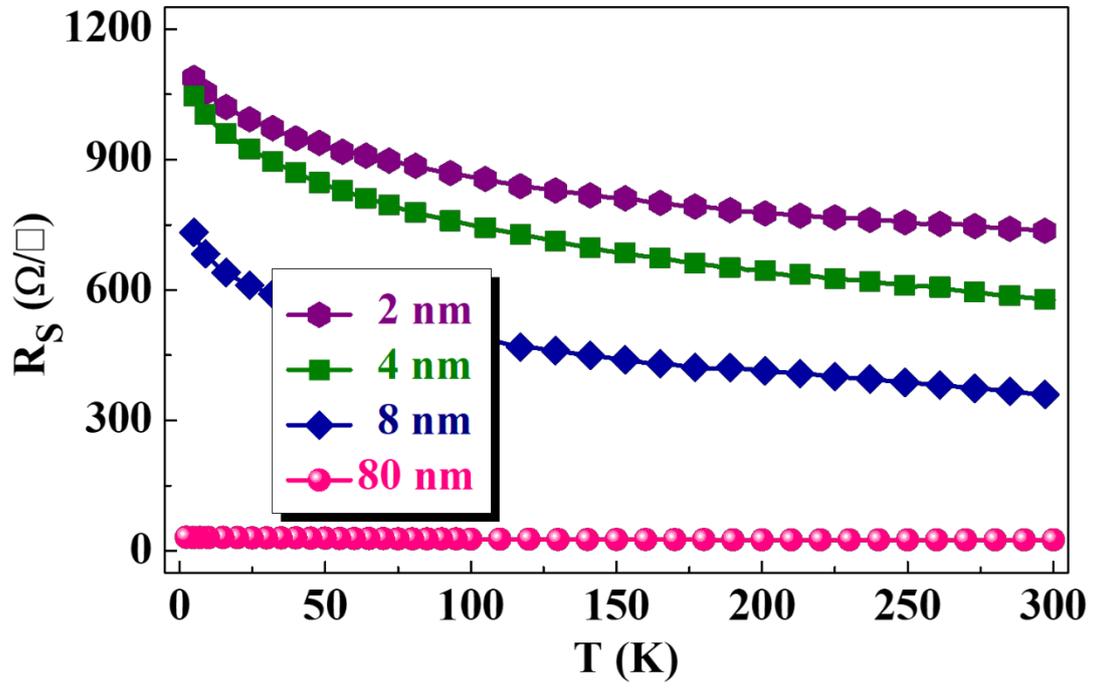

**(FIGURE-5)**

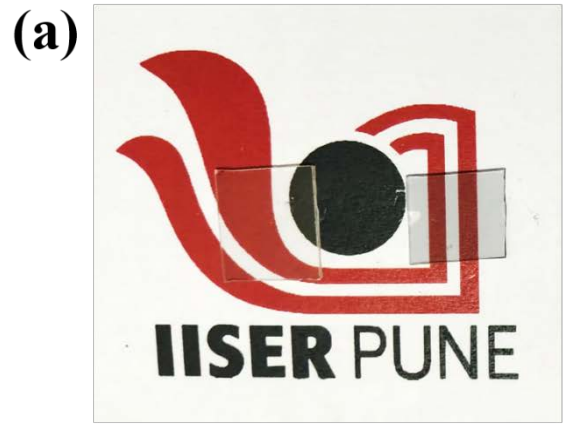

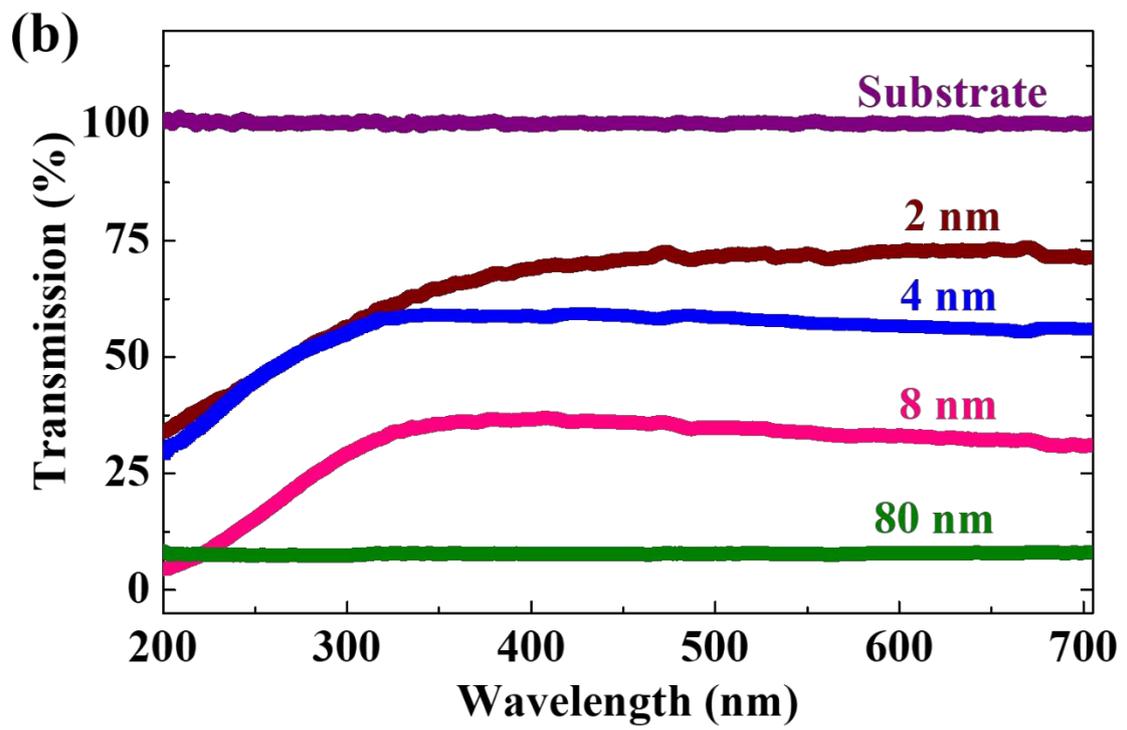

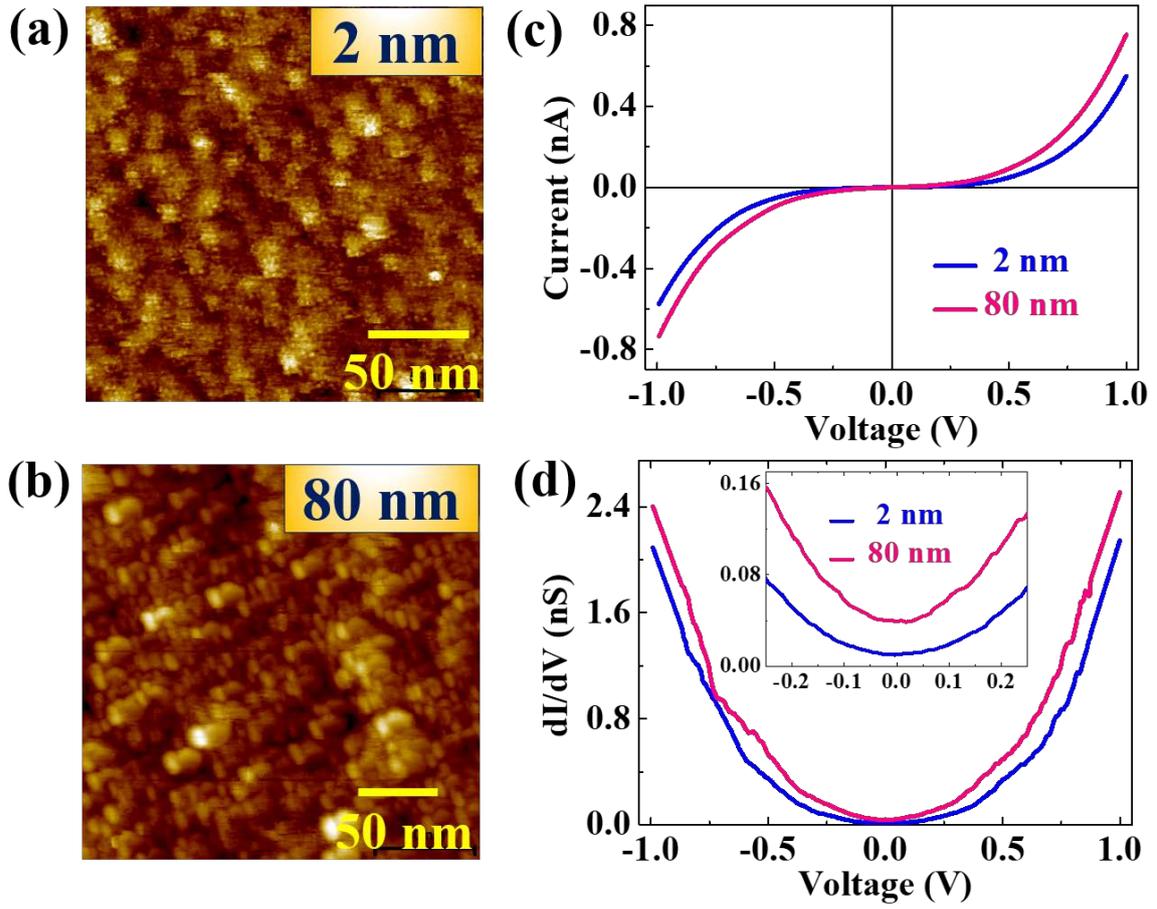

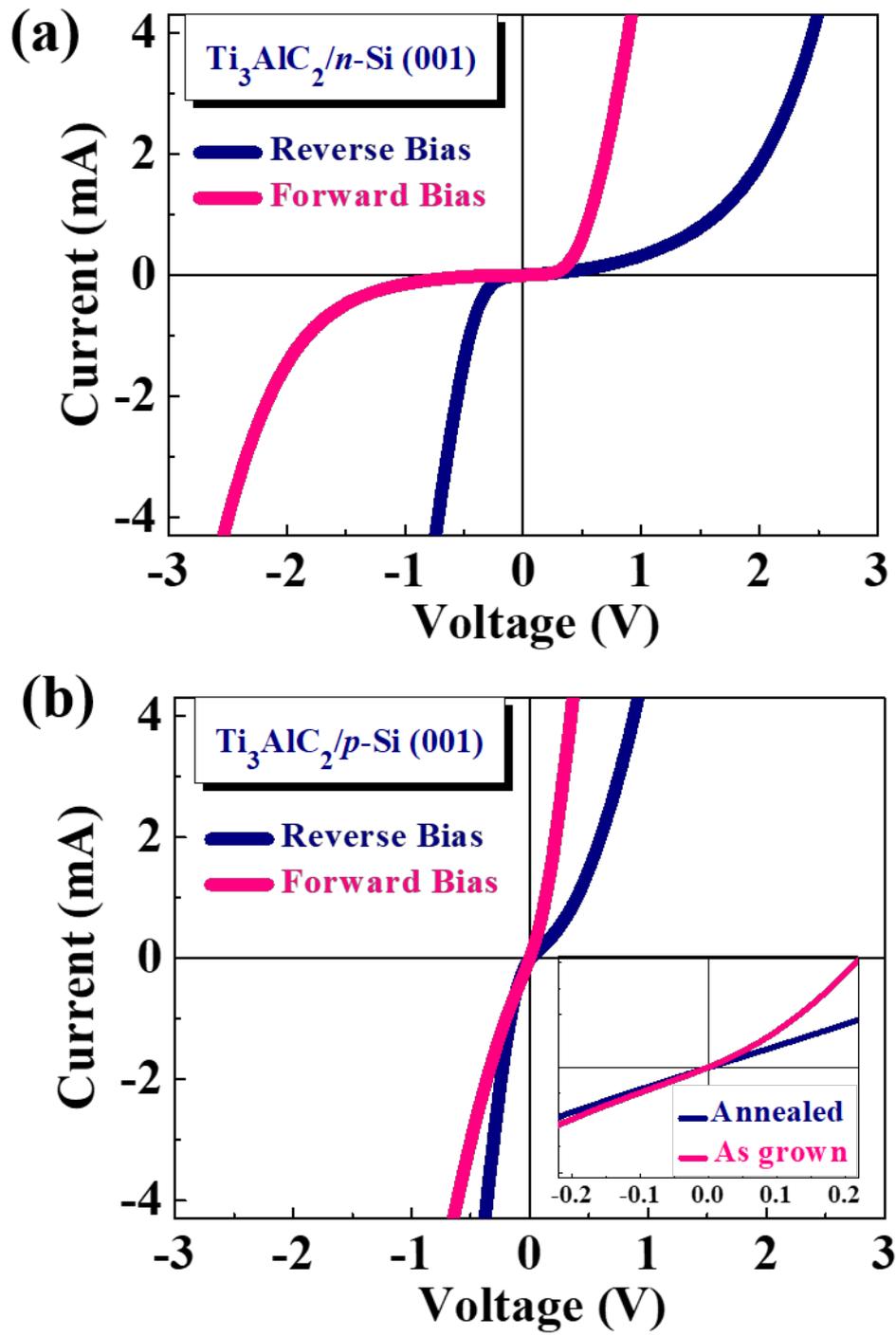

**(FIGURE-8)**

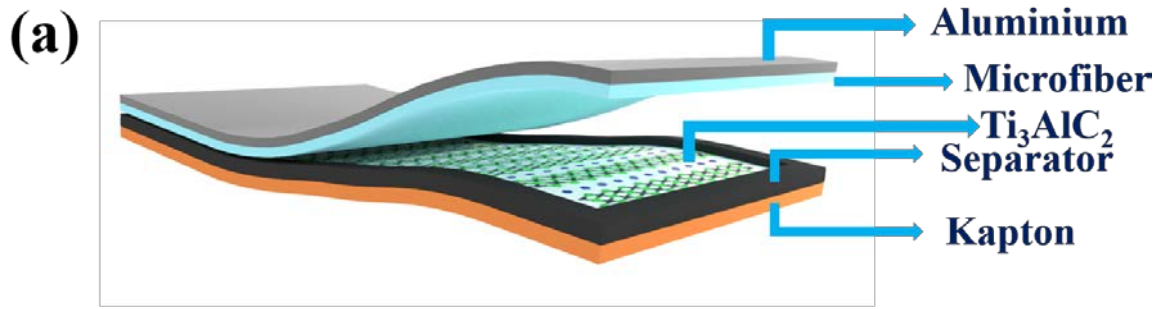

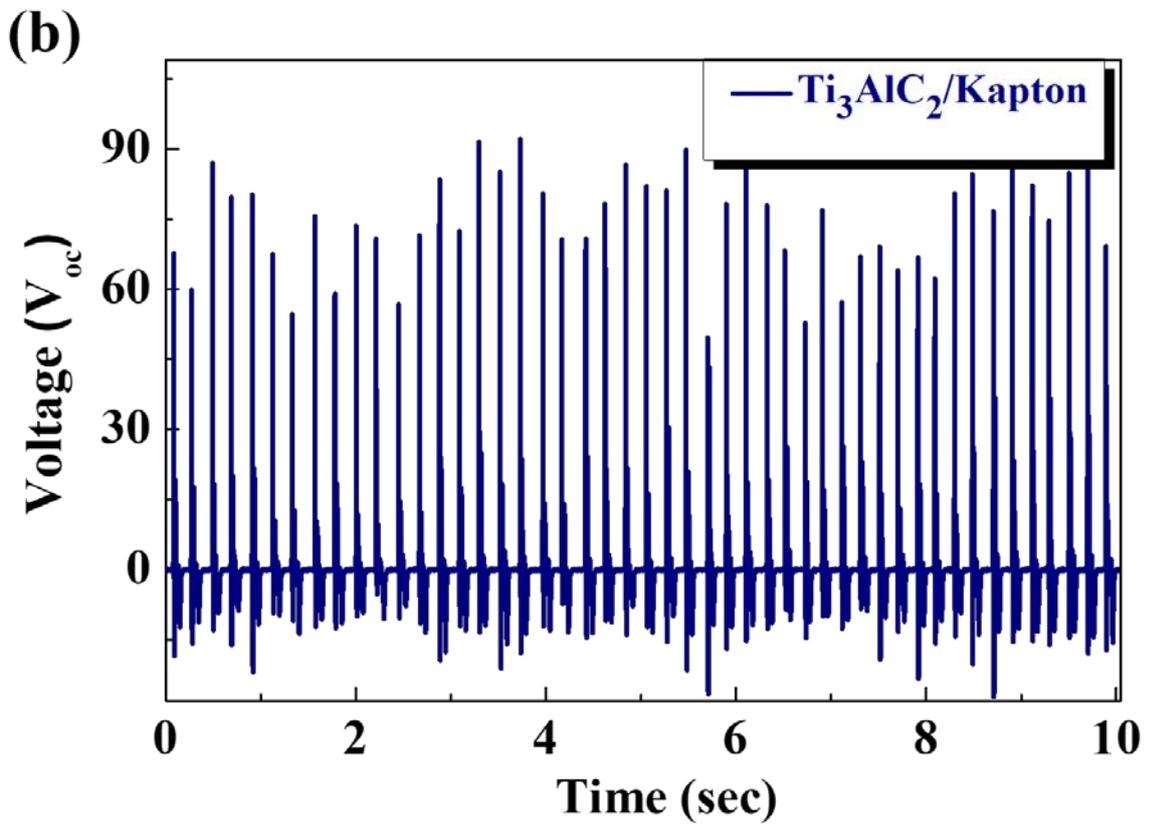

# Supporting Information

# Growth, Properties, and Applications of Pulsed Laser Deposited Nanolaminate Ti$_3$AlC$_2$ Thin Films

Abhijit Biswas[1], Arundhati Sengupta[1], Umashankar Rajput[1], Sachin Kumar Singh[1], Vivek Antad[2], Sk Mujaffar Hossain[1], Swati Parmar[1,3], Dibyata Rout[1], Aparna Deshpande[1], Sunil Nair[1] and Satishchandra Ogale[1,4]*

[1]*Department of Physics and Centre for Energy Science, Indian Institute of Science Education and Research (IISER), Pune, Maharashtra-411008, India*

[2]*Department of Physics, Nowrosjee Wadia College, Pune, Maharashtra 411001, India*

[3]*Department of Technology, Savitribai Phule Pune University, Pune, Maharashtra-411007, India*

[4]*Research Institute for Sustainable Energy (RISE), TCG Centres for Research and Education in Science and Technology (TCG-CREST), Kolkata-700091, India.*

**\*Correspondence:** satishogale@iiserpune.ac.in, satish.ogale@tcgcrest.org

**Keywords:** Ti$_3$AlC$_2$ phase, pulsed laser depositions, carrier transport, Schottky diode, Ohmic contact, triboelectric touch sensor

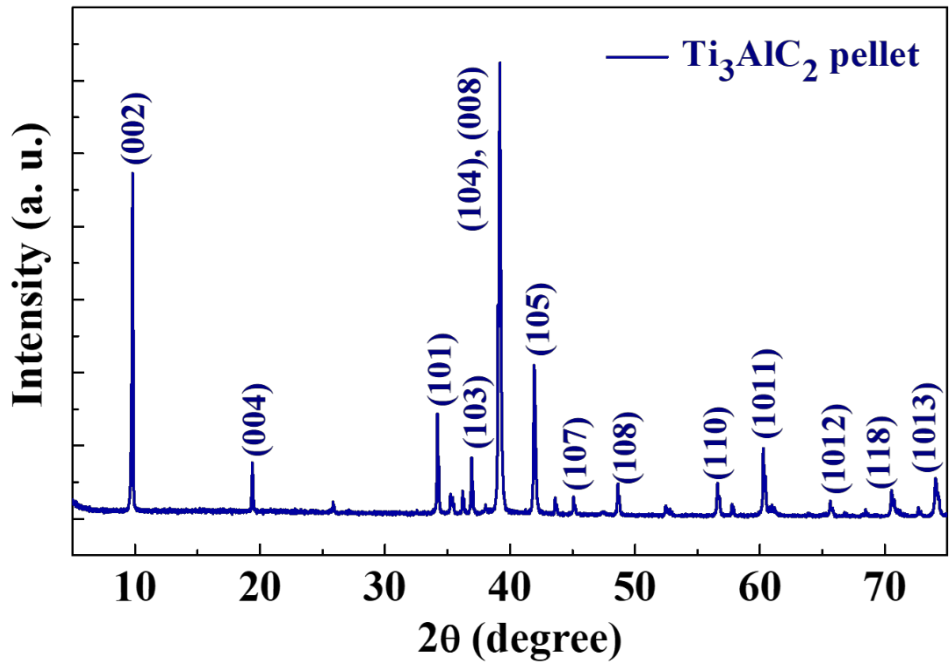

**FIG. S1.** X-ray diffraction of the prepared Ti$_3$AlC$_2$ powder showing the formation of single-phase material.

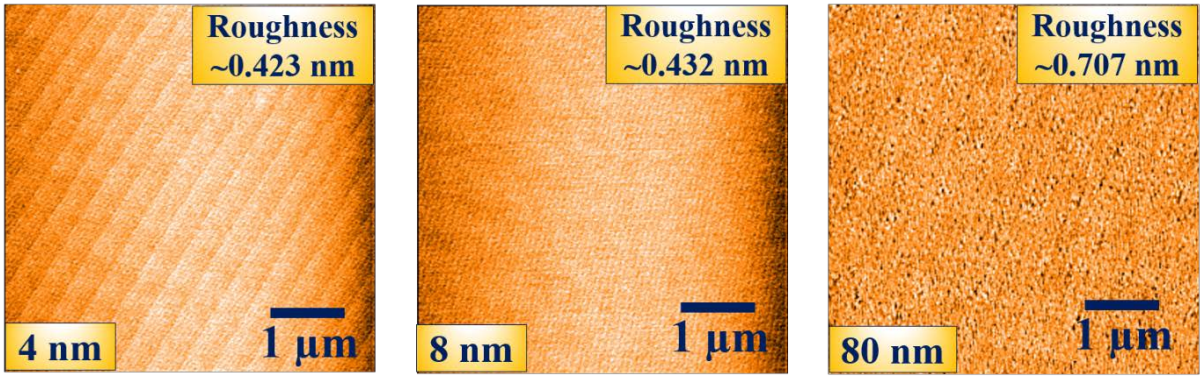

**FIG. S2.** Atomic force microscopy images of 4 nm, 8 nm and 80 nm Ti$_3$AlC$_2$ films on *c*-Al$_2$O$_3$ (0001) showing atomically flat surfaces with roughness less than one nm.

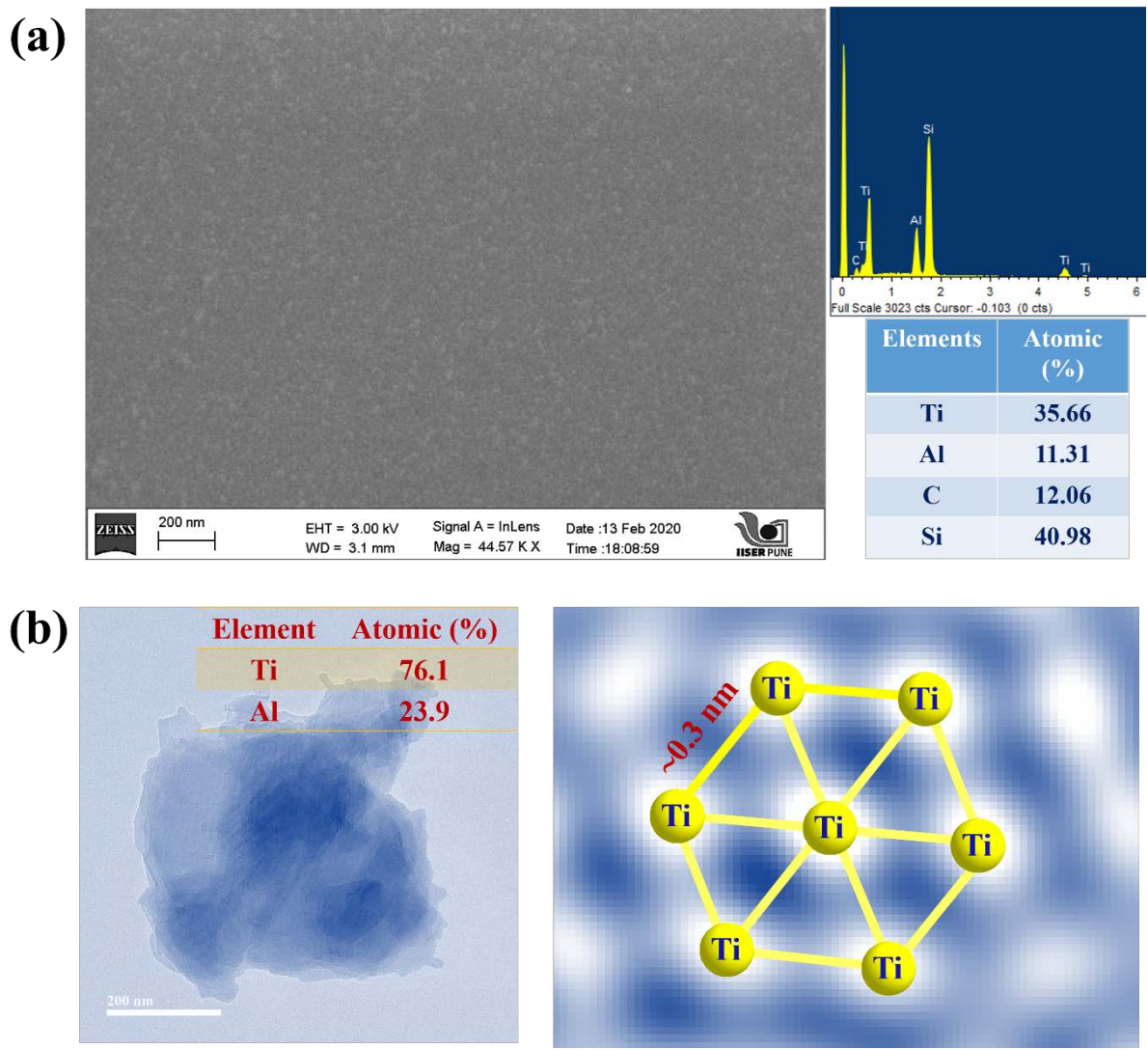

**FIG. S3.** (a) FESEM image and EDAX of $Ti_3AlC_2$ film grown on Si substrate (to avoid the contribution of Al from the $Al_2O_3$ substrate), also shows the atomic percentage of Ti and Al of ~3:1. (b) HRTEM images of $Ti_3AlC_2$ film showing the formation of hexagon (top view of $Ti_3AlC_2$). EDAX elemental analysis shows the atomic percentage of ~3:1 for Ti and Al.

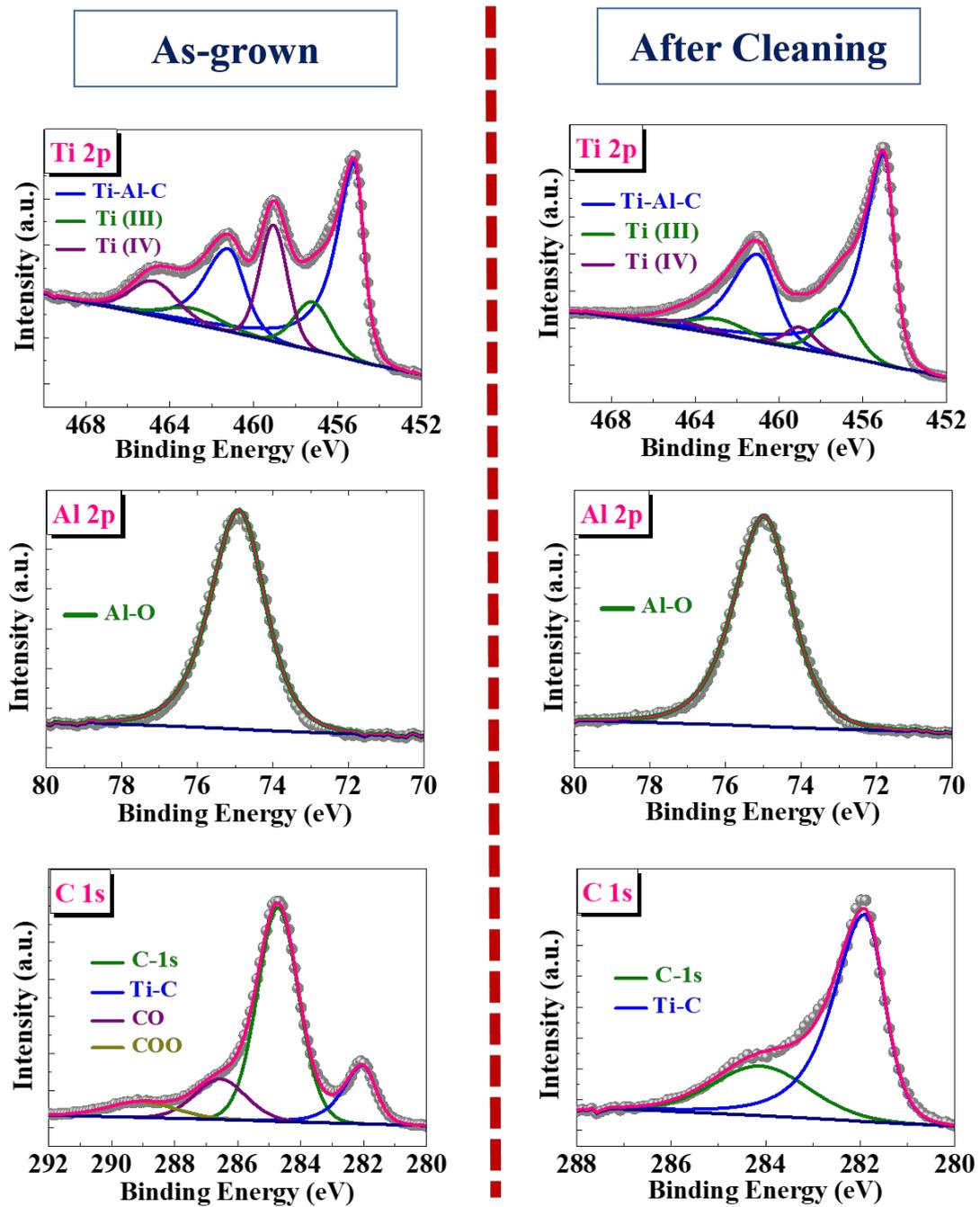

**FIG. S4.** XPS spectra of as grown and after in-situ sputter cleaned 80 nm $Ti_3AlC_2$ film.

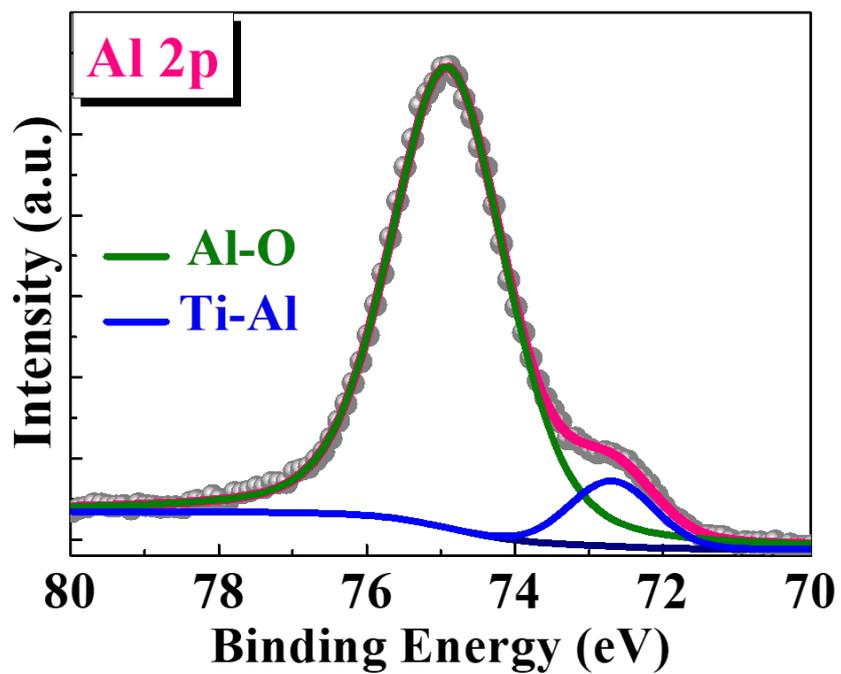

**FIG. S5.** Al-2p spectrum of $Ti_3AlC_2$ film grown at much lower growth temperature of 400 °C.

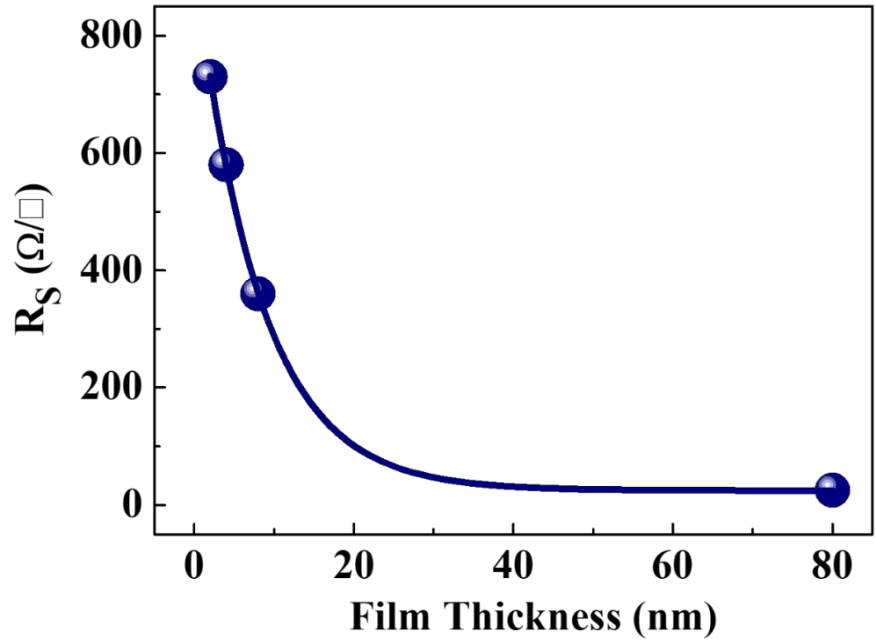

**FIG. S6.** Thickness dependent variation in sheet resistance of $Ti_3AlC_2$ films at room temperature.

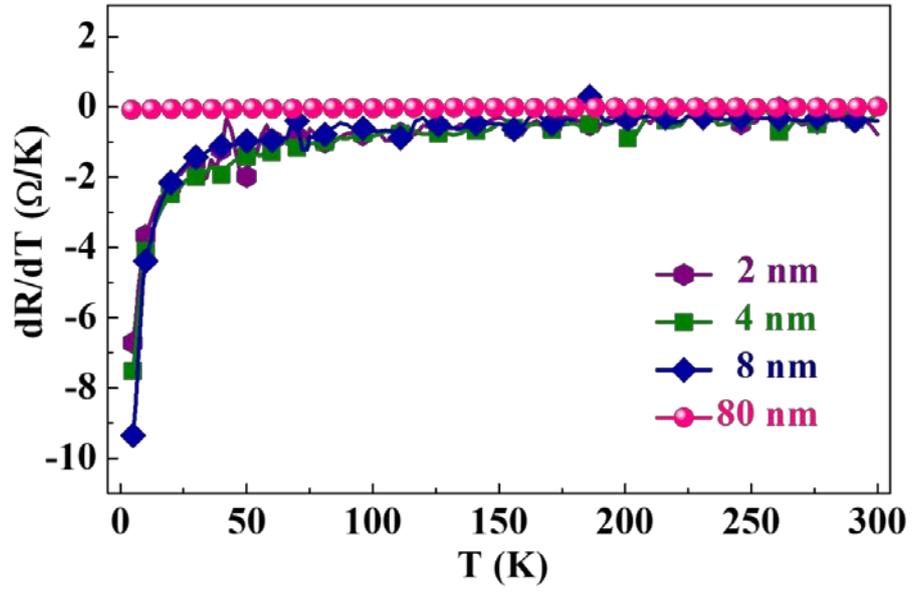

**FIG. S7.** Derivative of sheet resistance of $Ti_3AlC_2$ films.

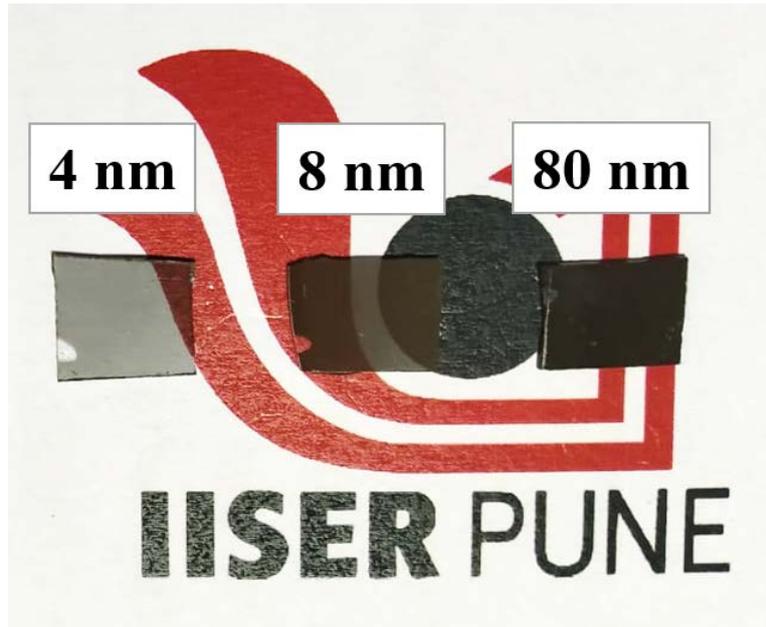

**FIG. S8.** Thickness dependent optical transparency of $Ti_3AlC_2$ films.

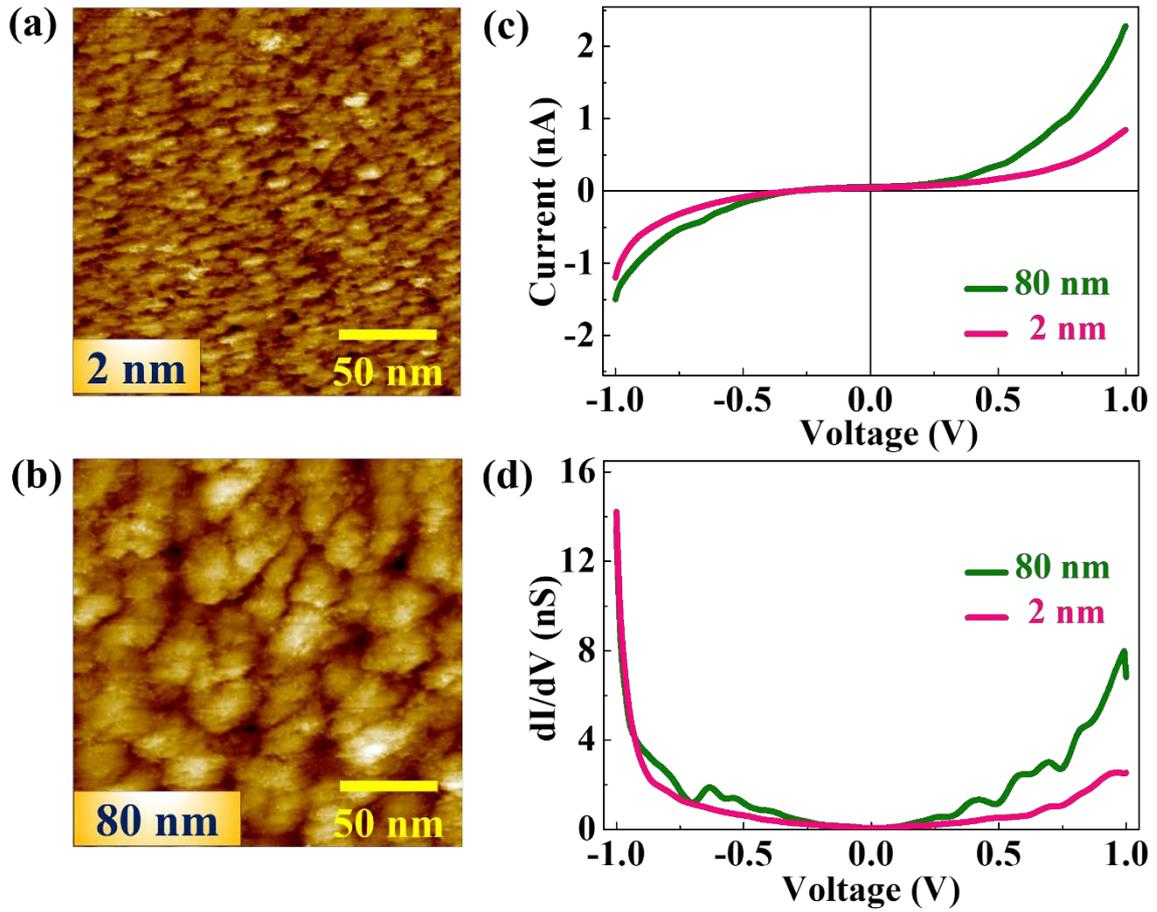

**FIG. S9.** (a)-(b) Scanning tunnelling microscopy (STM) topographic images, (c) current vs. voltage and (d) differential conductance (dI/dV) spectra of *as grown* 2 nm and 80 nm $Ti_3AlC_2$ films taken at room temperature within the applied bias voltage range of ±1V.

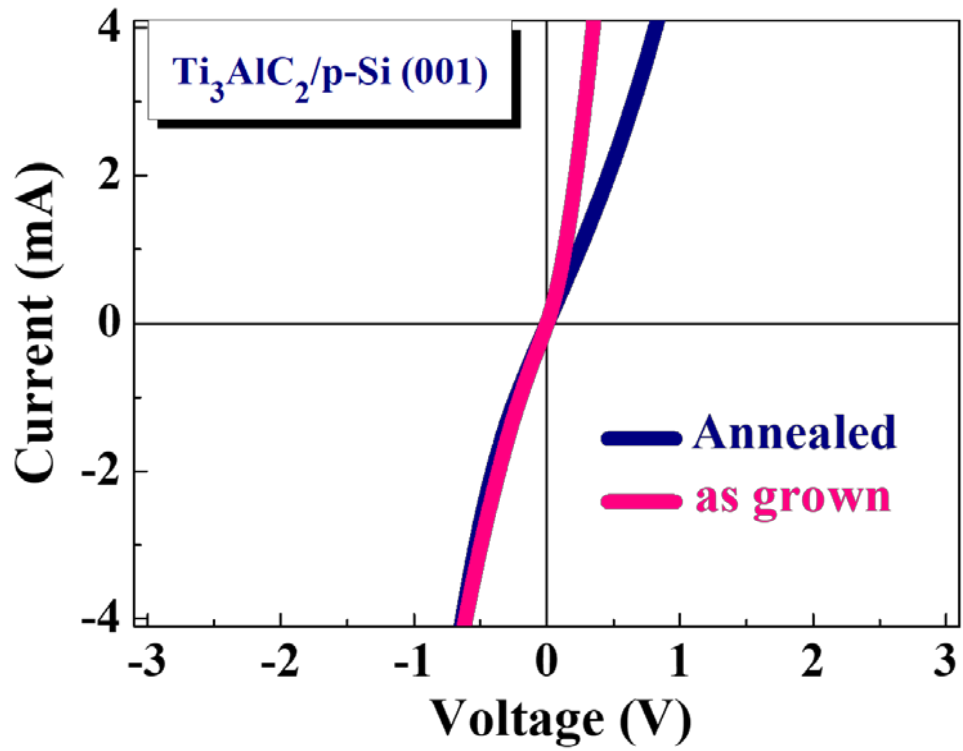

**FIG. S10.** Comparative I-V characteristics of Ti$_3$AlC$_2$/*p*-Si of as grown as well as film after in-situ annealing at 700 °C for 3 hrs.